# Ultrashort pulse generation from binary temporal phase modulation

**Anastasiia Sheveleva and Christophe Finot**

Laboratoire Interdisciplinaire CARNOT de Bourgogne,
UMR 6303 CNRS-Université Bourgogne-Franche-Comté,
9 Av. A. Savary, BP 47 870, 21078 DIJON Cedex, FRANCE

christophe.finot@u-bourgogne.fr

*Abstract*

We propose and numerically validate an all-optical scheme to generate a train of optical pulses. Modulation of a continuous wave with a periodic binary temporal phase pattern followed by a spectral phase shaping enables us to obtain ultrashort pulse trains. An ideal step phase profile as well as a profile arisen from a bandwidth-limited device are investigated. Analytical guidelines describing pulse trains formation and their characteristics are provided.

## 1. Introduction

The technological advances in the last decades have brought optical phase modulation technologies to the forefront of signal processing. Modern optical transmission systems benefit from these technologies in the field of optical networking.[1,2] Applications to the domain of ultrashort pulses at high-repetition rates have also stimulated great interest. Indeed, imprinting a quadratic spectral phase on a continuous wave that is phase modulated with a sinusoidal waveform in the temporal domain has been found efficient to generate ultrashort pulses at repetition rates of tens of GHz, [3,4] thus representing a convincing alternative approach to mode-locked laser sources [5] or to architectures based on nonlinear reshaping. [6,7] A versatility in the choice of a spectral phase profile opens up fruitful possibilities to tailor characteristics of pulse trains. As it has been recently experimentally demonstrated, the use of $\pi/2$ spectral phase shifts to replace the quadratic spectral phase profile[8,9] induces a noticeable decrease of the spurious background and of unwanted sidelobes enabling the generation of Gaussian Fourier-transform limited structures.

In this contribution, we propose to extend our study to new temporal profiles that can be used as an initial phase modulation. More precisely, instead of a sinusoidal waveform, we now numerically apply a periodic pattern of binary phase steps. Different spectral processing schemes are then tested and compared, such as a quadratic spectral phase, a triangular spectral phase or a Hilbert transform. Some analytical guidelines are derived for the case of an ideal pattern with a modulation depth of $\pi$. To get an idea of the temporal profiles that can be generated in a realistic case, numerical simulations taking into account the impact of the bandwidth limitation are also examined.

## 2. Ideal binary phase modulation

### 2.1 Initial phase ideal phase modulation

A sinusoidal pattern has found its use in most of the theoretical and experimental works exploiting periodic phase modulation techniques.[10] Indeed, this waveform makes less demands in terms of the RF bandwidth reducing a degree of distortion from the ideal pattern. However, with the progress of digital electronics, binary patterns have become routinely available. In this context and given the link that can be drawn between the temporal and spatial domains,[11,12] we can benefit from the knowledge gained in the field of diffractive optics where binary phase gratings have been the subject of many discussions.[13-15] We consider here a fully coherent continuous wave which is phase modulated by an ideal profile $\varphi(t)$ that is periodic (with a period $T_0$, leading to a frequency $f_0 = 1/T_0$) and infinite. Such a profile is plotted in Fig. 1(a) for two values of the phase offset $\Delta\varphi$ ($\pi$ and $\pi/4$, black and blue lines respectively). We limit our discussion to a duty-cycle of ½ so that the temporal phase modulation is defined over one period (i.e. between $-T_0/2$ and $T_0/2$) as :

$$\varphi(t) = \text{sgn}(-t) \; \Pi(t/T_0) \; \Delta\varphi/2 \tag{1}$$

where $\text{sgn}(t)$ and $\Pi(t)$ are the sign and the gate functions respectively.

The optical spectrum $s(f)$ of the phase-modulated signal is composed of equally spaced spectral components. Simple guidelines are available for a single cell

(i.e. one period), giving the complex spectral envelope of the following form :[14,15]

$$s(f) = T_0 \, \text{sinc}\left(\frac{\pi T_0}{2} f\right) \cos\left(\frac{\pi T_0}{2} f + \frac{\Delta\varphi}{2}\right) \quad (2)$$

The spectral intensity and phase profiles are illustrated in panels b and c of Fig. 1. These results provide a ground for discussion of several interesting points. In the case of a modulation depth of $\pi$, one can note that all the even components are suppressed, i.e. the spectral components are spaced by twice the frequency of the initial modulation. The amplitude of the $n^{th}$ component is given by $(n>0)$ [14] :

$$s\left((2n-1)f_0\right) = -\frac{2}{\pi} \frac{2}{2n-1} \sin\left(\frac{\Delta\varphi}{2}\right). \quad (3)$$

The ratio between the intensity of two successive unsuppressed (odd) spectral components is drawn by (see red mixed line) :

$$\frac{\left|s\left((2n+1)f_0\right)\right|^2}{\left|s\left((2n-1)f_0\right)\right|^2} = \left(\frac{2n-1}{2n+1}\right)^2. \quad (4)$$

Figure 1(c) outlines that the positive spectral components are phase-shifted by $\pi$ with respect to the negative components. When the modulation depth differs from $\pi$, we note that a central component appears in the spectrum (blue circles in Fig. 1(b2)). However, the overall shape of the spectrum remains identical and a similar spectral phase profile is observed. Let us also remark that a binary intensity modulation will lead to the same overall spectrum structure.

## *2.2 Spectral phase processing*

We now consider the impact of various spectral processing schemes on the temporal intensity profile. The first profile that we consider is a quadratic spectral phase that typically arises from the dispersion. The same kind of spectral phase profile appears when diffraction takes place. So following the space/time analogy, [11,12] the dispersion is the temporal counterpart of the diffraction. As the phase modulation is not limited in time, we have to consider the near field regime typical of Fresnel diffraction [16] where the various temporal orders do not get temporally isolated. [17] The dispersive propagation of the periodic pattern will therefore exhibit a behavior representative of the Talbot carpet observed in the temporal domain, [18] as illustrated in Fig. 2 for an initial phase offset of $\pi$. The quadratic phase profile is $\exp(-i\,\beta_2\,z\,\omega^2/2)$, with $\omega$ being the angular frequency and $\beta_2$ the second-order dispersion coefficient of a device of length $z$. The linear medium of propagation can be a dispersive fiber or a fiber Bragg grating. [19] Results can then be normalized by the Talbot length $z_T$ defined by :

$$z_T = \frac{1}{\pi\,|\beta_2|\,(\Delta f)^2}, \quad (5)$$

where $\Delta f$ is the frequency spacing between two non-zero spectral components ($\Delta f = 2f_0$ in our ideal case). After a propagation distance of $z_T/4$, the phase modulation is converted into a binary intensity modulation (red curve, Fig. 2b), typical of a Talbot array illuminator. [20-22] After $z_T/2$, a continuous wave is reconstructed as expected from the self-imaging process. The structure with the highest peak power is obtained after a propagation distance of $0.19\,z_T$ and is plotted with a black line in Fig. 2(b). However, its non-monotonic waveform is less appealing in terms of practical applications.

Another approach in spectral processing of the binary phase modulated wave relies on applying a triangular spectral phase profile. The slope of the linear spectral profile is chosen so as to impart a $\pi/2$ phase change on every $f_0$ spectral component. Experimentally such treatment can be applied using phase programmable spectral shapers [23] or fiber Bragg gratings. [24] A temporal profile obtained after the spectral processing is influenced by the initial modulation depth of the binary pattern as shown in Fig. 3. The phase modulation is efficiently converted into ultrashort temporal structures equally spaced by $T_0/2$. Heavy tails and a difference in the intensity profile are present at any initial modulation depths except for the case of $\Delta\varphi = \pi$ where the repetition rate is doubled.

Inspired by the spectral phase profile seen in Fig. 1(c), the third processing we have investigated was the photonic multiplication of the spectrum by the sign function, i.e. applying a $\pi$ phase shift between the negative and positive spectral components while cancelling the central component. Such a spectral treatment corresponds to a Hilbert transform. [25,26] Hilbert transform has been initially demonstrated in spatial free-space optics [27] but now it can be all-optically performed by a spectral programmable filter, in Bragg gratings, [28,29] in a SOI microdisk chip [30] or in photonic crystal nanocavity. [31] An alternative technique that may provide similar results is the use of optical differentiation. [32,33] The temporal intensity profiles arisen after the spectral processing are summarized in Fig. 4 based on which we can note that the repetition rate is doubled whatever the initial phase offset is. The resulting intensity profile plotted on a logarithmic scale in Fig. 4(b) confirms that, with the suppression of the central component, the shape obtained after the Hilbert transform is not influenced by the value of the initial phase depth .

## *2.3 Analytical insights*

In order to get further insight into the waveform that is achieved after the Hilbert transform, one may consider the Hilbert transform $H(t)$ of a single cell (with a phase profile provided by Eq. (1)). For $\Delta\varphi = \pi$ the analytical expression takes the following form:

$$H(t) = -\frac{i}{\pi} \ln\left(\frac{t^2}{\left|t^2 - (T_0/2)^2\right|}\right) \qquad (6)$$

Results are plotted in Fig. 5 on a linear and logarithmic scales. The resulting waveform is Fourier-transform limited and can also provide a first approximation of the shape achieved after dispersive propagation (see Fig. 2b, blue dotted line). The central part of the highly peaked waveform is characterized by a diverging behavior at $t = 0$. We can also remark that the Hilbert transform of a single cell does not perfectly fit the profile obtained for the periodic train. Indeed, $H(t)$ spans over a temporal duration that largely exceed a single period so that it will influence the neighboring pulses. Results obtained considering a series of 7 cells (blue line) are in a good agreement with the profiles arisen from an ideal and infinite train (red line).

## *3. Temporal phase modulation with bandwidth limitation*

The discussion we developed in the previous section assumes an ideal phase jump. Though this assumption enables us to easily derive interesting analytical guidelines, for more realistic predictions it is required to take into account the finite bandwidth of the phase modulation that will limit the steepness of the transition between the two phase levels in the binary profile. As a first approximation, we consider that the bandwidth limitations can be modelled by Gaussian filter with a full width at half maximum of $14\,f_0$. The resulting phase profile is shown in Fig. 6(a) (black line) for a modulation depth of $\pi$. Even if the steepness has been reduced compared to the ideal case, the sharpness of the edges remains well above a standard temporal sinusoidal phase modulation having the same modulation depth (red line). The resulting optical spectrum is given in panel (b) and is compared with spectra originated from the sinusoidal phase modulation (red line) and the ideal envelope predicted by Eq. (2) (dashed black line). We can note that the bandwidth limitation leads to the emergence of even components in the optical spectrum. The spectral extend obtained from a binary pattern is significantly higher than the one resulting from a sinusoidal modulation where the typical ratio between the optical intensity $s_s$ of spectral components is defined as [34]:

$$\left|\frac{s_S\left((n+1)f_0\right)}{s_S(n\,f_0)}\right|^2 = \frac{(\Delta\varphi/2)^2}{4(n+1)^2}. \qquad (7)$$

The temporal profiles arisen from the dispersive propagation of a periodically modulated wave with a phase depth of $\Delta\varphi = \pi$ are depicted in Fig. 7 with respect to the propagation distance normalized by the Talbot length (Eq. (5) with $\Delta f = f_0$). Pulse structures obtained after $z = 0.054\,z_T$ are plotted in panel (b) and have a fwhm duration of only $0.1450\ T_0$. The waveform is strongly impaired by a residual background that contains a non-negligible portion of the energy, therefore limiting the peak power compared to what could be expected from a Fourier-transform limited structure (dotted blue line). Results obtained with a triangular spectral phase shaping inserting a $\pi/2$ phase shift between two successive spectral components are shown in Fig. 8(a1). In order to improve the result, the central component of the modulated spectrum has been suppressed. Due to the bandwidth limitation the phase modulation depth of the binary pattern is no longer $2\pi$ periodic regarding $\Delta\varphi$. A severe background also emerges between two pulses spaced by $T_0/2$. However, for a modulation depth of $5.4$ rad, the level of these sidelobes becomes acceptable as shown in panel (b) of Fig. 8 (blue line). The fwhm temporal duration is $0.07\ T_0$ and the peak power has been increased by a factor higher than $9$ with respect to the initial continuous wave. The emergence of these ultrashort structure is consistent with the qualitative conclusions reported in the work of R. N. Shakhmuratov.[35] The duration that is achieved depends on the steepness of the phase drop and longer pulses are obtained for a stronger bandwidth limitation.

When a Hilbert transform is applied (panel a2 of Fig. 8), a doubling of the repetition rate is recovered. The pulses are however significantly longer with a duty cycle of only $0.32$ and a peak power increase limited to a factor $2.5$ (see black line of Fig 8(b)).

## *4. Conclusions*

To conclude, we have discussed the patterns achieved after a spectral phase processing of a continuous wave periodically modulated with a binary temporal phase. Whereas the quadratic spectral phase leads to a phase-to-intensity conversion known as the Talbot array illuminator, other spectral phase modulation schemes are found even more interesting, such as a triangular profile or the Hilbert transform. This last scheme has been found efficient to double the repetition rate and to achieve ultrashort structures whatever the initial modulation depth is. Analytical insight into the pulse waveform has been described. When non-ideal binary phase modulation is taken into account, the main qualitative properties of the resulting pattern are preserved and one can expect to generate pulse train pattern at the doubled-up repetition rate using a photonic Hilbert transform. A triangular spectral profile combined with the suppression of the central component has also enabled the generation of ultrashort pulses with increased peak-power. To consider a realistic scenario we have investigated temporal waveforms obtained after taking into account a Gaussian-like bandwidth limitation of the initial binary phase profile. In addition to that we have explored other linear filters such as a Butterworth frequency response which have driven to the same qualitative conclusions. In the presented study we have focused on a duty cycle of $1/2$ for the initial binary phase modulation. Other values of this factor could provide another degree of freedom to be explored. [14]

Our numerical results demonstrate that binary phase modulation associated with a convenient spectral phase processing can be potentially involved in various applications such as ultrashort pulse generation [9], optical sampling [36] or noiseless application scheme. [37] We can also anticipate that the proposed scheme could be combined with an initial phase modulation which frequency linearly varies over time. Thereby pulse trains with jitters in the pulse-to-pulse delays and temporal widths could be achieved.[38]

### *Acknowledgements:*

We acknowledge the support of the Institut Universitaire de France (IUF), the French Investissements d'Avenir program and the Agence Nationale de la Recherche, France (Graduate School EIPHI ANR-17-EURE-0002). We thank Ugo Andral, Julien Fatome, Bertrand Kibler and Sonia Boscolo for fruitful discussions.

**Figure captions:**

Fig. 1 Comparison of the temporal and spectral properties obtained for phase modulation depths $\Delta\varphi$ of $\pi$ and $\pi/4$ (black and blue color, respectively) (a) Temporal phase profile applied over a continuous wave. The vertical red dashed lines delimit the elementary ceil we consider for Eq. (1). (b) Optical power spectrum $|s|^2$. The spectrum of the periodic waveform (circles) is compared to the shape of the spectral envelope predicted by Eq. (2) and (4) (dashed lines and mixed red lines, respectively). (c) Spectral phase profile.

Fig. 2 Temporal intensity profiles achieved for different levels of quadratic spectral phase. (a) Longitudinal evolution of the temporal profile according to the propagation length in the dispersive element. (b) Details of the intensity profile obtained at a propagation distance of $0.25\ z_T$ (red line) and $0.19\ z_T$ (black line). The blue dashed line corresponds to the Fourier-transform limited pulse train.

Fig. 3 (a) Evolution of temporal intensity profile according to the level of the initial phase modulation depth $\Delta\varphi$ after a triangular spectral phase has been applied to the modulated signal. (b) Details of the intensity profile obtained for $\Delta\varphi = \pi/4$, $\pi/2$ and $\pi$ (blue, red, and black lines respectively).

Fig. 4 (a) Evolution of temporal intensity profile according to the level of the initial phase modulation depth $\Delta\varphi$ after a Hilbert transform (b) Details of the intensity profile obtained $\Delta\varphi = \pi/4$, $\pi/2$ and $\pi$ (blue, red and black lines respectively).

Fig. 5 Temporal intensity profile after a Hilbert transform of the phase modulated signal plotted on (a) linear and (b) logarithmic scales. Results obtained from Eq. (6) based on a single elementary cell (black line) are compared with the result taking into account 7 cells (blue line) and the profile arisen from an infinite periodic modulation (red line).

Fig. 6 Comparison of the temporal and spectral properties obtained for periodic modulation with the phase depth $\Delta\varphi$ of $\pi$ impaired by a finite frequency bandwidth (black) and for a sinusoidal phase modulation (red). (a) Temporal phase profile. The results are compared to the ideal modulation considered in section 2 (dotted blue line). (b) Optical power spectrum $|s|^2$. The spectrum of the binary phase modulated wave (black circles) is compared to the shape of the spectral envelope predicted by Eq. (2) (black dashed lines). The spectrum obtained from a sinusoidally modulated wave is compared with the spectral intensity ratio given by Eq. (7). (c) Spectral phase profile.

Fig. 7 Temporal intensity profiles achieved for different levels of quadratic spectral phase in the case of a temporal phase modulation with a limited bandwidth. (a) Longitudinal evolution of the temporal profile according to the propagation length in the dispersive element. (b) Details of the intensity profile obtained at a propagation distance of $0.054$ $z_T$ (black line). The blue dashed line corresponds to the Fourier-transform limited pulse train.

Fig. 8 (a) Evolution of temporal intensity profile according to the level of the initial phase modulation depth $\Delta\varphi$ after a spectral processing based on a triangular phase and suppression of the central component (a1) or after a Hilbert transform (a2). (b) Details of the intensity profile obtained for $\Delta\varphi = 5.4$ rad (triangular spectral phase and suppression of the central spectral component, blue line) and $\Delta\varphi = \pi$ (Hilbert transform, black line).

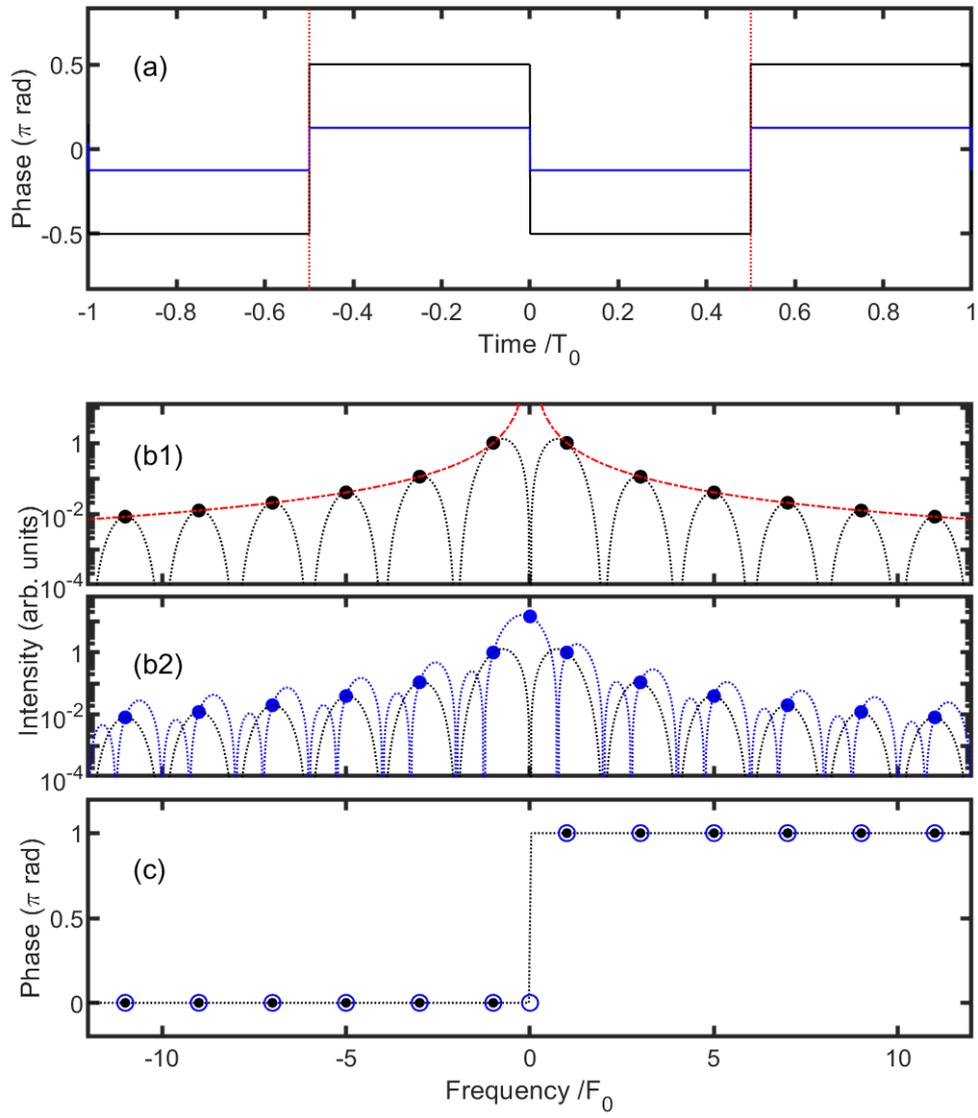

Figure 2

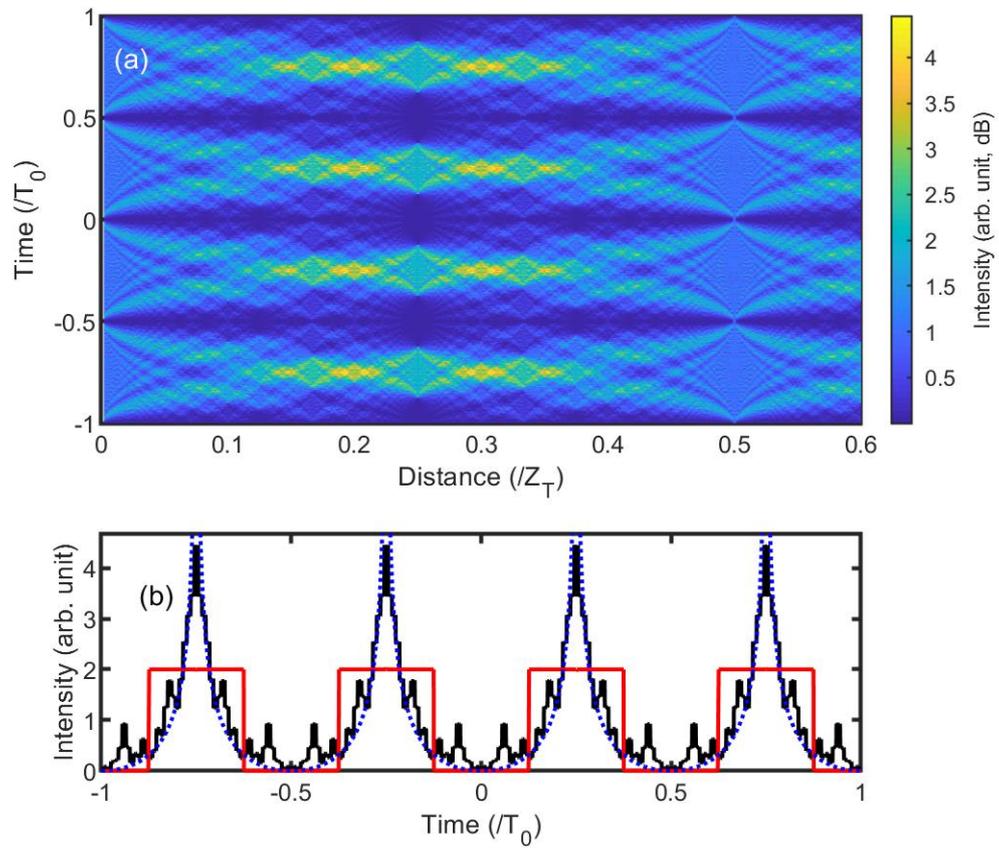

Figure 3

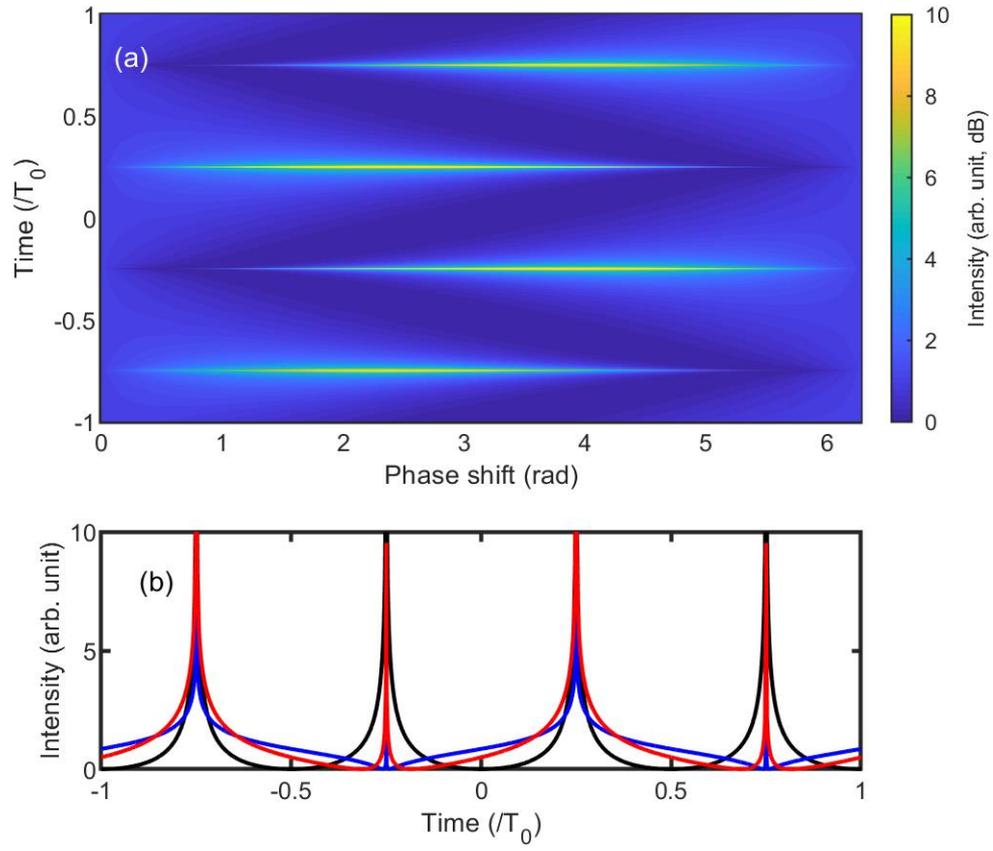

Figure 4

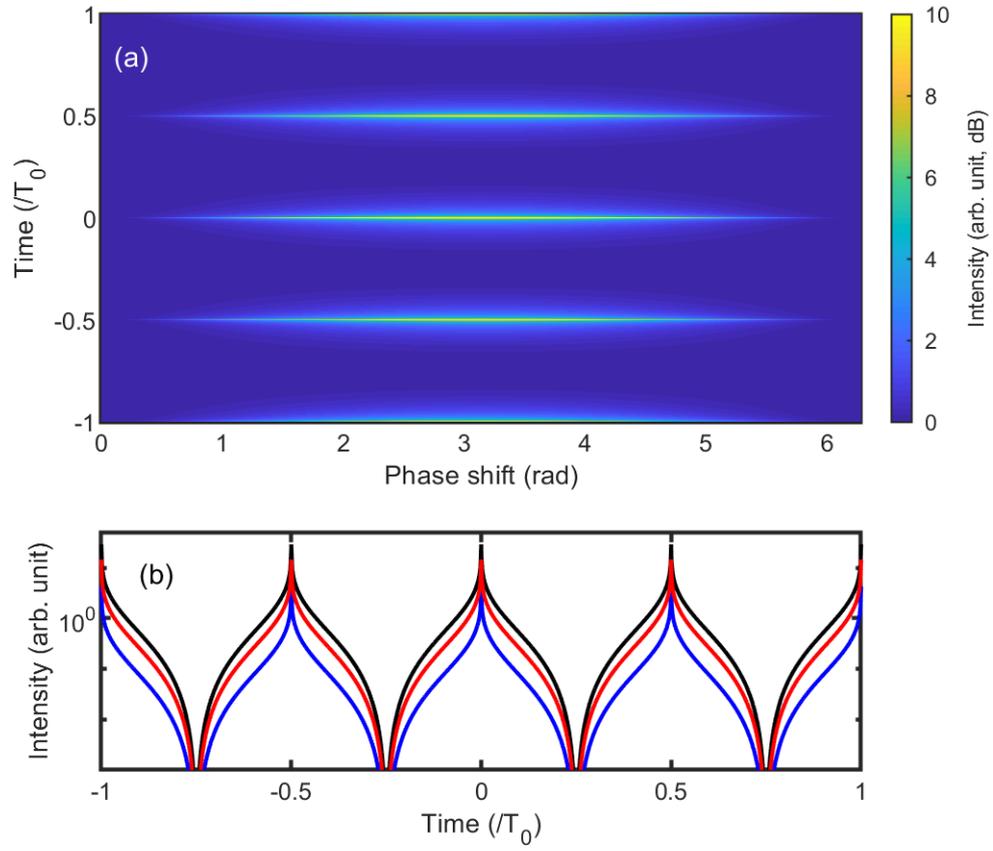

Figure 5

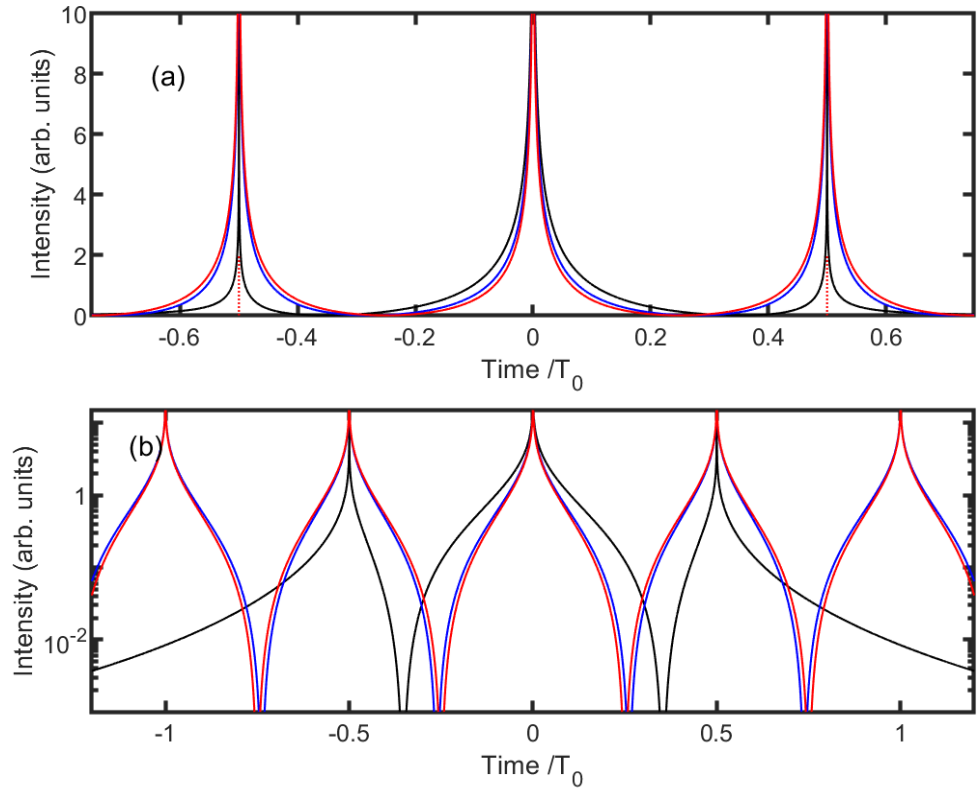

Figure 6

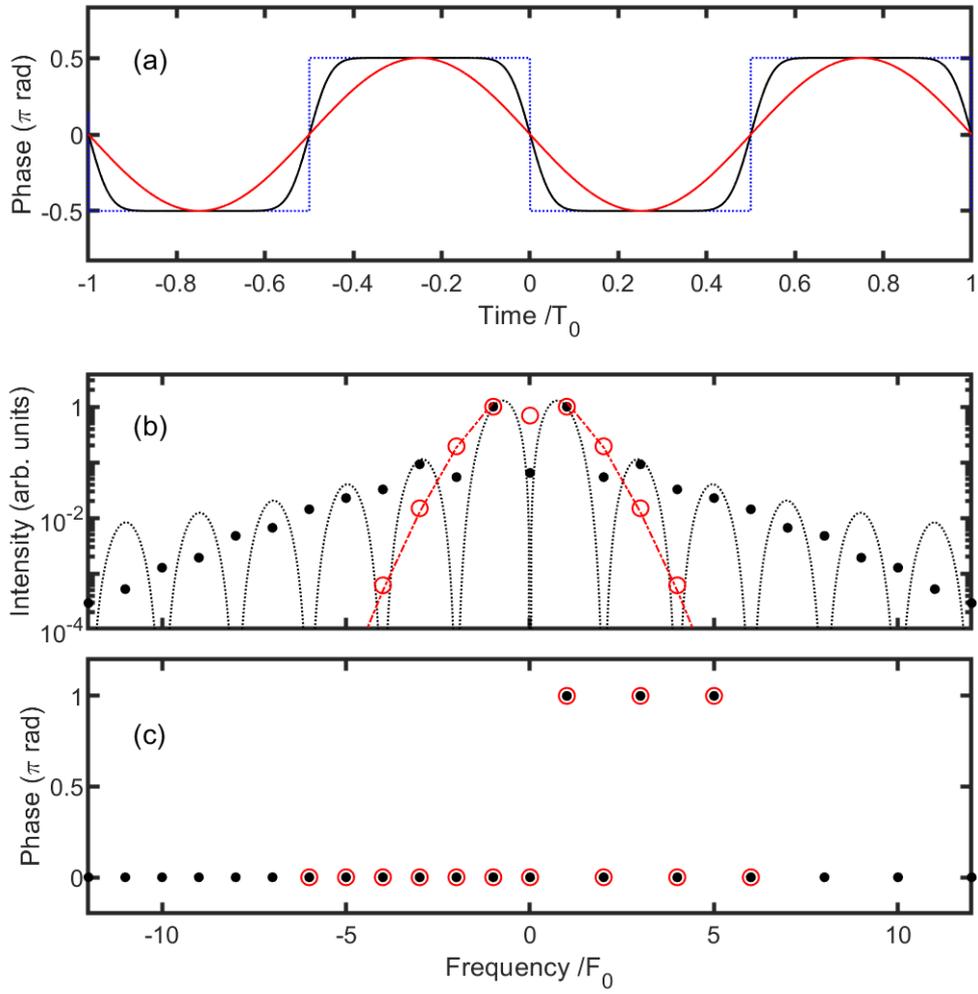

Figure 7

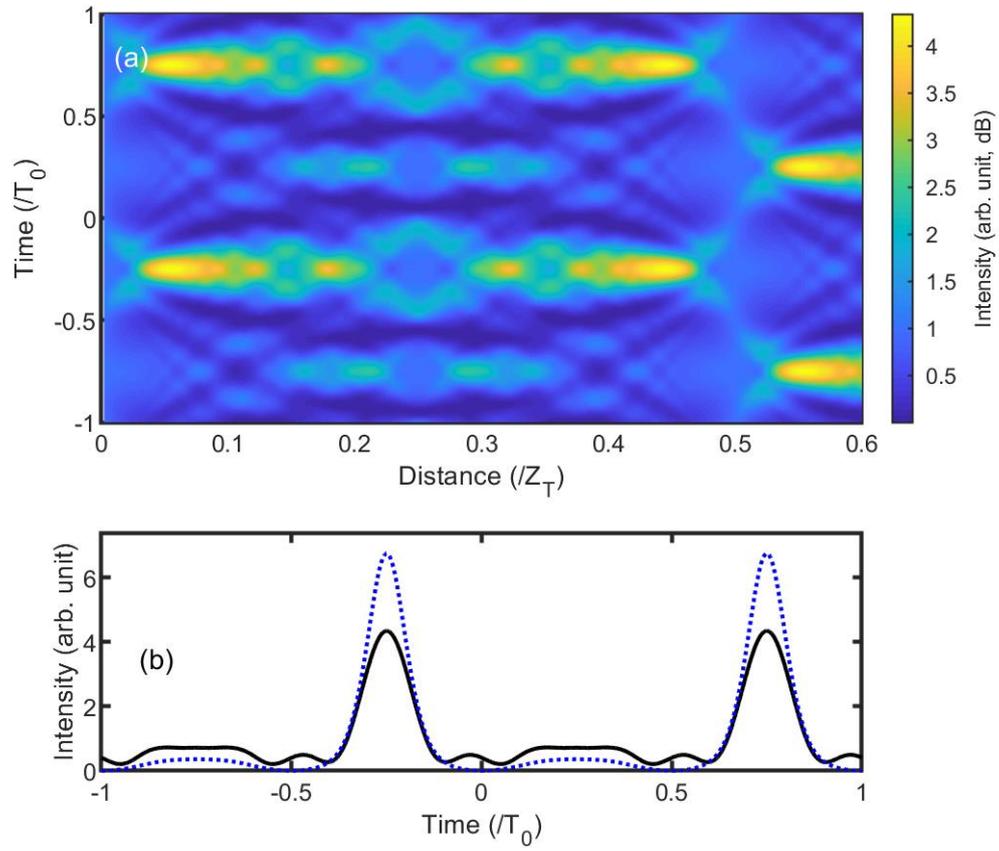

Figure 8

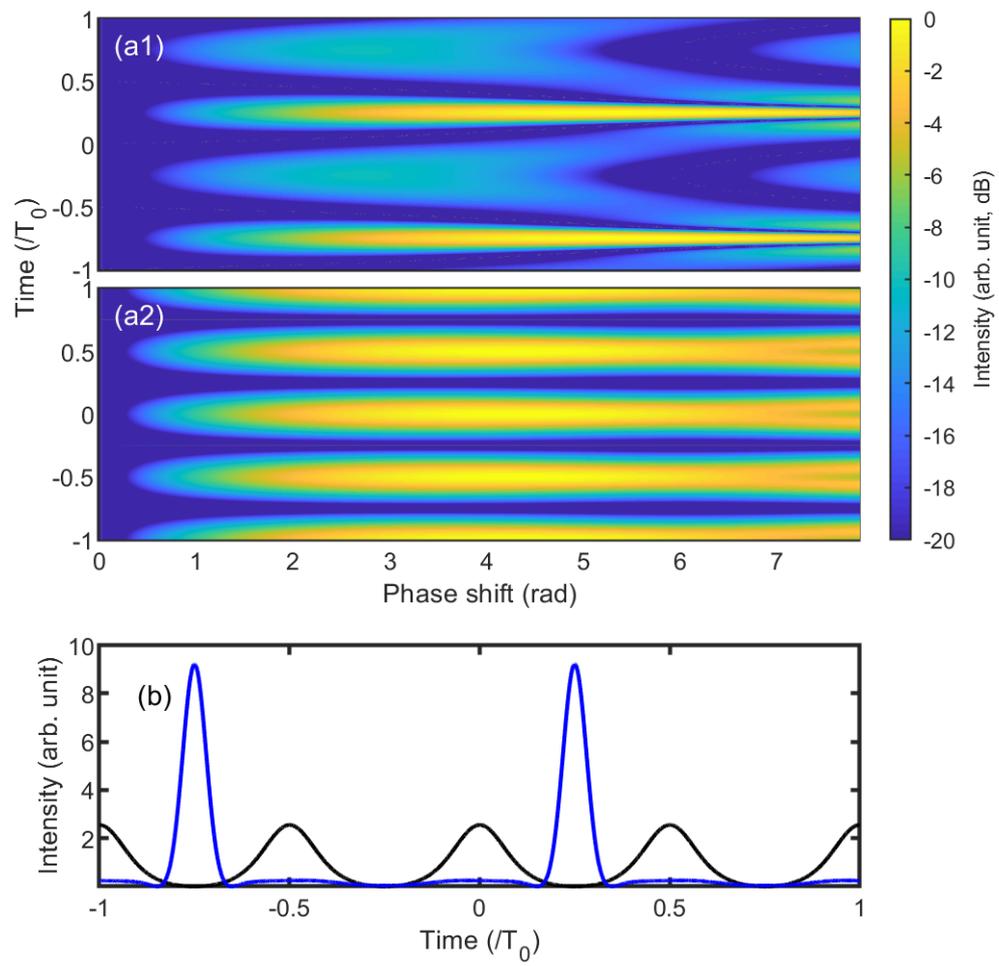